\documentclass[10pt]{article}
\usepackage{amsmath}
\usepackage{amssymb}
\usepackage{graphicx}
\usepackage{cite}
\usepackage{color}
\usepackage[labelfont=bf,labelsep=period,justification=raggedright]{caption}

\makeatletter
\renewcommand{\@biblabel}[1]{\quad#1.}
\makeatother
\begin{document}

\date{November 26, 2012}

\begin{flushleft}
{\Large
\textbf{A thermokinetic approach to radiative heat transfer at the nanoscale}
}
\\
Agust\'{\i}n P\'{e}rez-Madrid$^{1,\ast}$, 
Luciano C. Lapas$^{2,\ast}$,
J. Miguel Rub\'{\i}$^{1,\ast}$
\\
\bf{1} Departament de F\'{\i}sica Fonamental, Facultat de F\'{\i}sica, Universitat de Barcelona, Mart\'{\i} i Franqu\`{e}s 1, Barcelona 08028, Catalunya, Spain
\\
\bf{2} Universidade Federal da Integra\c{c}\~{a}o Latino-Americana, Caixa Postal 2067, Foz do Igua\c{c}u 85867-970, Paran\'{a}, Brazil
\\
$\ast$ To whom correspondence should be addressed. (APM) E-mail: agustiperezmadrid@ub.edu. (LCL) E-mail: luciano.lapas@pq.cnpq.br. (JMR) E-mail: mrubi@ub.edu. 
\end{flushleft}


\section*{Abstract}

Radiative heat exchange at the nanoscale presents a challenge for several areas due to its scope and nature. Here, we provide a thermokinetic description of microscale radiative energy transfer including phonon-photon coupling manifested through a non-Debye relaxation behavior. We show that a lognormal-like distribution of modes of relaxation accounts for this non-Debye relaxation behavior leading to the thermal conductance. We also discuss the validity of the fluctuation-dissipation theorem. The general expression for the thermal conductance we obtain fits existing experimental results with remarkable accuracy. Accordingly, our approach offers an overall explanation of radiative energy transfer through micrometric gaps regardless of geometrical configurations and distances.


\section*{Introduction}

There is a general consensus that the search for clean sources of energy with no climatic and environmental impact constitutes a major strategic objective at present. In this sense, nanoscale thermal radiation conversion offers a source for intensive clean energy generation. Thermal radiation has always been an active field of study. Throughout the 19$^{\text{th}}$ century, great scientists (Boltzmann, Stefan, Rayleigh, etc.) dedicated a considerable effort to this problem which was completely solved at the turn of the 20$^{\text{th}}$ century due to Planck's contribution to the founding of quantum mechanics. After this, it seemed to be a well-established fact that the maximum power extracted from a hot body depended on the temperature as $T^{4}$.

However, recently, thanks to modern technological advances, it has been shown that energy exchange through thermal radiation at nanometric distances breaks by several orders of magnitude the limits posed by Stefan-Boltzmann law for black body radiation. Moreover, near-field thermal radiation is approximately monochromatic and reveals itself coherent in space and time, which may lead to stationary interference phenomena in a micro-cavity. Therefore, this monochromaticity and coherence along with the overcoming of Stefan-Boltzmann limits, all of these distinguishing features together confer near-field radiation a great potential for future applications in nanotechnology and, as we have said at the beginning, energy conversion as well. Several reviews on this issue have been written recently, see Ref.~\cite{Zhang07, Biehs12, Basu07} by way of example.

An ever-increasing number of investigators has marked the recent history of the research on near-field radiation. Polder and van Hove~\cite{Polder71} first studied heat transfer between two objects at nanometric scales maintained at different temperatures by following a stochastic or fluctuational electrodynamics formalism established by Rytov et al.~\cite{Rytov59}. Recently, it has been emphasized~\cite{Mulet02} that surfaces modes as included in the solution of Maxwell's equations in Ref~\cite{Polder71} can greatly enhance the heat flow. Experiments showing that the heat flow at the nanoscale is indeed greater than the blackbody radiation limit among materials supporting surface modes were reported in Ref~\cite{Rousseau09,Shen09} and between two gold surfaces in Ref~\cite{Shen12}. Likewise, Pendry~\cite{Pendry99} gave a simple derivation of the expression found by Polder and van Hove and some interpretation in terms of heat transfer channels in addition to a discussion of the maximum heat flux. The problem was reformulated by using a Landauer-like approach by Biehs et al.~\cite{Biehs10}. Finally, Sasihithlu and Narayanaswamy~\cite{Sasihithlu11} performed a discussion of the proximity approximation. All these approaches have something in common: the linear response regime and, consequently, the fluctuation-dissipation theorem (FDT), whose validity is not guaranteed at this level. Thus, an approach based on the contributions of fluctuating dipole effects seems to be the heart of a considerably simplified treatment of energy transfer at the nanoscale~\cite{Shen09,Rousseau09,Domingues05}. These approaches include a wide range of phenomena in which the energy between molecules is dominated by dipole-dipole interactions, also known as the F\"{o}rster energy transfer~\cite{Forster48}.

Nonetheless, as two nanostructures thermalized at different temperatures come closer to each other, the distribution of charges and currents becomes asymmetric and therefore, defies description in terms of dipolar interactions. Hence, it becomes clear that one must bear in mind higher order effects beyond the dipole~\cite{Madrid08} and also include other contributions to thermal conductance quite common in disordered amorphous materials, leading to a generalization of the FDT mentioned above~\cite{Madrid09}.

In this article, we will go deeper into these aspects by showing how phonon-photon coupling effects account for microscale radiative energy transfer by considering non-Debye relaxation due to the excess of modes in the low and high frequencies in the bulk material~\cite{Madrid08,Madrid09,Madrid10}. In the current literature on disordered systems, the non-Debye refers to an excess of modes of vibration over the Debye level observed in inelastic light (Raman) and neutron scattering. From here, we obtain a general expression for the heat transfer coefficient including both Debye and non-Debye contributions, providing an overall explanation of the energy transfer through micrometric gaps. The findings of our theory fit existing experimental results with a high degree of accuracy.


\section*{Methods}

Our theoretical framework has been described in previous publications~\cite{Madrid09,Madrid10} and is briefly summarized here. We consider a gas of quanta distributed in phase space according to the probability density $\rho(\boldsymbol{\Gamma },t)$, where $\boldsymbol{\Gamma }=(p,x)$, $p\equiv \left\vert \mathbf{p}\right\vert $, and $x\equiv \left\vert \mathbf{x}\right\vert $, with $\mathbf{p}$ and $\mathbf{x}$ being the momentum and position of a quanta, respectively. Here, we must pay attention to the fact that $\rho (\boldsymbol{\Gamma },t)$ possesses dimensions of $h^{-3}$, $h$ being Planck's constant. The thermodynamic description that we propose entails the formulation of the second law of thermodynamics, which can be carried out by means of the Gibbs entropy postulate~\cite{deGroot84}, 
\begin{equation}
S(t)=-k_{B}\int \rho (\boldsymbol{\Gamma },t)\ln \frac{\rho (\boldsymbol{\Gamma },t)}{\rho _{eq}(\boldsymbol{\Gamma })}d\mathbf{p}d\mathbf{x}+S_{eq}. \label{entropy1}
\end{equation}
This equation gives us the nonequilibrium entropy of the gas of quanta plus the bath, with $S_{eq}$ being the equilibrium entropy and $\rho_{eq}$ the equilibrium probability density.

In general, entropy is produced due to irreversible processes, in such a manner that irreversible processes in nonequilibrium systems are described by means of currents, thermodynamic forces (affinities), and the entropy production rate, which is always positive. It is precisely this positive character of the entropy production that enables us to derive relaxation equations. Since the probability density is conserved the existence of a density gradient $\partial \rho (\boldsymbol{\Gamma },t)/\partial \boldsymbol{\Gamma }=\left( \partial \rho /\partial x,\partial \rho /\partial p\right) $ yields a current $\mathbf{J}\left( \boldsymbol{\Gamma },t\right) =(J_{x},J_{p})$ which unleashes a relaxation process $\partial\rho /\partial t=-\partial \mathbf{J}/\partial \boldsymbol{\Gamma }$. In addition, this current satisfies the relation $\mathbf{J}\left( \boldsymbol{\Gamma },t\right) =-\mathbf{D}(\rho )\cdot \partial \rho (\boldsymbol{\Gamma },t)/\partial \boldsymbol{\Gamma }$, derived from the entropy production, which can be obtained from Eq. (\ref{entropy1})~\cite{Madrid09,deGroot84}. Here, $\mathbf{D}(\rho )$ is a material-dependent quantity, the matrix of diffusion coefficients, satisfying Onsager's symmetry principle. In general, due to the tensorial character of the diffusion matrix, both currents $J_{x}$ and $J_{p}$ are coupled
\begin{equation}
J_{\alpha }=-D_{\alpha \beta }\frac{\partial \rho }{\partial \beta} \text{; with $\alpha,\beta =x,p$,}  \label{currents}
\end{equation}%
where $D_{\alpha \beta }$ are the diffusion matrix components.

\begin{figure}[!ht]
\begin{center}
\includegraphics[width=4in]{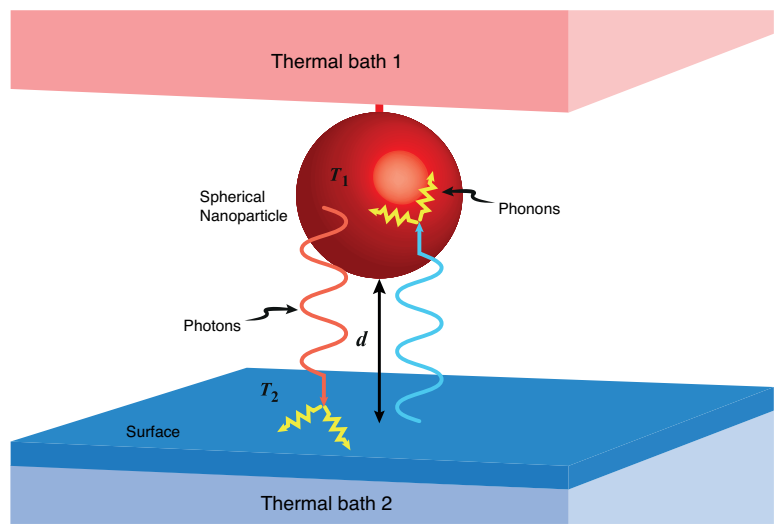}
\end{center}
\caption{ \textbf{Schematic diagram of the radiation exchanged via conventional radiative transfer.} Elastic collision of photons with atoms or
molecules of materials and phonon-photon coupling contributions between a sphere and a plate maintained at different temperatures, $T_{1}$ and $T_{2}$, separated by a distance $d$.} \label{fig:scheme}
\end{figure}

The main contribution of this article is to apply this formalism to the description of the radiative heat transfer between a nanosphere and a plate at different temperatures, $T_{1}$ (hot) and $T_{2}$ (cold), separated by a distance $d$ (see Fig. \ref{fig:scheme}). We assume that heat transfer results from two different mechanisms. (I) On one hand, we consider a conventional radiative heat exchange involving the dynamics of quasiparticles as the result of two simultaneous processes: elastic emission and absorption of hot photons from the medium at $T_{1}$ and elastic emission and absorption of cold photons from the medium at $T_{2}$; these processes also involve the presence of surface modes~\cite{Pendry99,Mulet02}. (II) The second mechanism behind heat transfer has to do with the excitation of coupled resonant modes from the collection of acoustic states related to defective soft structures in a disordered regime~\cite{Denisov90,Shintani08,Chumakov11}. This process encompasses inelastic scattering of the impinging radiation, which is linked to nonequilibrium contributions. In both scenarios, there is no diffusion in configuration space since quanta are massless particles~\cite{Madrid09}, and as a consequence $J_{x}=0$, which brings about $J\left(\boldsymbol{\Gamma },t\right) \longrightarrow J_{p}\left( \boldsymbol{\Gamma },t\right) $. Hence, from Eq. (\ref{currents}) one obtains the appropriate diffusion current in momentum space 
\begin{equation}
J_{p}\left( \boldsymbol{\Gamma },t\right) =\left( \frac{D_{pp}D_{xx}}{D_{xp}} -D_{px}\right) \frac{\partial \rho \left( \boldsymbol{\Gamma},t\right)}{\partial x} \equiv -\frac{\hbar }{\tau (\rho )}\frac{\partial \rho ( \boldsymbol{\Gamma },t)}{\partial x}\text{,}  \label{rate}
\end{equation}
where $\tau (\rho )$ is the relaxation time, which depends on the diffusion matrix components.

\subsection*{Near-field analysis}
In the near-field regime, confinement of the electromagnetic waves in a micrometric gap separating neighboring nanostructures introduces peculiar effects in the spectrum of the thermal radiation. Here is where the collective modes (phonons) excited in the material by the impinging radiation come into play. To discern whether confinement effects are important or absent we shall assume a cut-off wavelength, the thermal wavelength of a photon $\lambda _{T}=hc/k_{B}T$, which is proportional to the Wien's displacement law through a proportionality constant, i.e. $\lambda _{T}=\beta /T$, where here $\beta \approx 2.82143937212$. Actually, when $d\gg \lambda _{T}$ we have a blackbody spectrum of radiation. One may wonder what happens when $d\lesssim \lambda _{T}$. Since according to Heisenberg's principle $\bigtriangleup x\bigtriangleup p\geqq h$, assuming that the maximum value of $\bigtriangleup x$ is $d$, one obtains $\bigtriangleup p\geqq h/d$, and thus the minimum value of $\bigtriangleup p$ is $h/d$, leading to $\bigtriangleup E\geqq h\varepsilon c/d$, with $0<\varepsilon <1$. Here, $\varepsilon $ is, precisely, the inverse of the refractive index. Hence, not all frequencies are possible and the frequency of resonance $\omega _{R}=2\pi \varepsilon c/d$ appears.

In these circumstances, integrating by parts the resultant continuity equation gives us 
\begin{equation}
\int_{-\infty }^{p}\left(-\frac{\partial \rho }{\partial t}\right)dp^{\prime}= J_{p}
\end{equation}
with $J_{p}\sim p/th^{3}$, where we have assumed that $J_{p}\left(\boldsymbol{\Gamma },t\right) =0$ at $p=-\infty $. Performing a second integration of Eq. (\ref{rate}) through from 1 to 2, we find the net current 
\begin{equation}
J(p,t)=-\frac{\hbar }{\tau ^{\ast }(t)}\left( \rho _{2}-\rho _{1}\right)\text{,} \label{rate2}
\end{equation}
where $\rho _{j}(p,t)=\rho \left( p,x=x_{j},t\right) $ is related to the population of quasiparticles at $x_{j}$ and $\rho _{1}+\rho _{2}=1$. Moreover, $J(p,t)\equiv \left( 1/\tau ^{\ast }\right) \int_{1}^{2}\tau (\rho)J_{p}dx$, with $\tau ^{\ast }(t)=\int \rho \tau (\rho )dpdx$ corresponding to a hierarchy of relaxation times ubiquitous in complex systems.

In the stationary state, once the system has thermalized at temperatures $T_{1}$ and $T_{2}$, $\rho _{j}(p,t)\longrightarrow 2N(\omega,T_{j})/h^{3}$; the factor $2$ comes from the polarization of a photon and $N(\omega ,T)$ being the averaged number of quasiparticles in a elementary cell of volume $h^{3}$ of the phase-space given by Planck's distribution~\cite{Planck91}, $N(\omega ,T)=1/\left[ \exp \left( \hbar\omega /kT\right) -1\right] $. Besides, $\tau ^{\ast }(t)\longrightarrow \tau ^{\ast }(\omega )$; therefore leading to a stationary value, i.e. $J(p,t)\longrightarrow J_{st}(\omega )$. Thus, the heat flow $Q$ can be obtained from the sum of all the contributions as 
\begin{equation}
Q=\int \varepsilon cJ_{st}(\omega )d\mathbf{p}=\frac{1}{2}h^{3}\varepsilon c\int_{\omega _{R}}^{\infty }J_{st}(\omega )g\left( \omega \right) d\omega \text{,} \label{heat}
\end{equation}
where $\mathbf{p}=\left( \hbar \omega /\varepsilon c\right) \boldsymbol{\Omega }_{p}$, with $\boldsymbol{\Omega }_{p}$ being the unit vector in the direction of $\mathbf{p}$, and the distribution of frequencies is given by 
\begin{equation}
g(\omega )=\frac{\omega ^{2}}{\pi ^{2}(\varepsilon c)^{3}}\text{.} \label{DOS}
\end{equation}

At this point, it is worthwhile to make a short digression about the physical meaning of the time scale $\tau ^{\ast }\left( \omega \right) $. It has been known for a long time~\cite{Yager36} that for most condensed systems in time-dependent fields, the orientation polarization behavior can, as a good approximation, be characterized by a relaxation time distribution ($\tau ^{\ast }\left( t\right) $); this behavior is generally meant as dielectric relaxation. In harmonic fields, this implies that the complex dielectric permittivity in the frequency range corresponding to the characteristic times for the molecular reorientation can be written as 
\begin{equation}
\epsilon \left( \omega \right) =\epsilon _{\infty }+\left( \epsilon_{s}-\epsilon _{\infty }\right) \int_{0}^{\infty }\frac{\zeta (\tau )}{1+i\omega \tau }d\tau \text{,} \label{dielectric_function}
\end{equation}
where $\epsilon _{s}$ is the static dielectric constant and $\epsilon_{\infty }$ is the permittivity at the infinite frequency. Here, the relaxation time distribution $\zeta \left( t\right) $ satisfies the normalization condition, $\int_{0}^{\infty }\zeta \left( \tau \right) d\tau=1$ and Eq. (\ref{dielectric_function}) constitutes a generalization of the Debye treatment based on Clausius-Mossotti equation~\cite{Debye29} 
\begin{equation}
\epsilon \left( \omega \right) =\epsilon _{\infty }+\left( \epsilon_{s}-\epsilon _{\infty }\right) \frac{1}{1+i\omega \tau }\text{.} \label{debye equation}
\end{equation}
This Debye's equation, Eq. (\ref{debye equation}), follows after Fourier-transforming the relaxation function $\phi \left( t\right) =\exp\left( -t/\tau \right) $ which coincides with the normalized dielectric function $\left( \epsilon _{s}-\epsilon \left( t\right) \right) /\left(\epsilon _{s}-\epsilon _{\infty }\right) $. Note that if we assume a time-independent time scale $\tau $ in Eq. (\ref{rate2}), after integration we shall obtain the Debye relaxation function in terms of $\exp \left(-t/\tau \right) $. Hence, we can understand $\tau ^{\ast }\left( \omega\right) $ as defined through the relation 
\begin{equation}
\int_{0}^{\infty }\frac{\zeta \left( t\right) }{1+i\omega \tau }d\tau \equiv \frac{1}{1+i\omega \tau ^{\ast }\left( \omega \right) } \label{generalized time}
\end{equation}
which shows that the Callen-Welton FDT~\cite{Callen51} is not valid at this level. In fact, unlike here, the FDT is related to decaying equilibrium fluctuations characterized, precisely, by a single relaxation time.

Now, let's return to the main topic after that brief digression. Note that according to Eq. (\ref{heat}), in the limit $d\rightarrow \infty $, 
\begin{equation}
Q=\frac{1}{2}h^{3}\varepsilon c\int_{0}^{\infty }J_{st}(\omega )g\left(\omega \right) d\omega , \label{heat_blackbody}
\end{equation}
giving us the blackbody radiation limit provided $\tau ^{\ast }(\omega)^{-1}\propto \omega $, $Q\propto \left[ \left( k_{B}T_{1}\right)^{4}-\left( k_{B}T_{2}\right) ^{4}\right] $. On the other hand, in the limit $d\rightarrow 0$, $Q\rightarrow 0$ which in contrast to the descriptions based on evanescent surface waves avoids divergences in the heat flux in a self-consistent way. Nonetheless, for finite $d$ we can rephrase the expression of the heat current given by Eq. (\ref{heat}) introducing a new varible $\omega =1/s$ 
\begin{eqnarray}
Q &=&\frac{1}{2}h^{3}\varepsilon c\int_{0}^{\omega _{R}^{-1}}J_{st}\left[\omega \left( s\right) \right] g\left[ \omega (s)\right] \frac{ds}{s^{2}} \notag \\
&\simeq &\frac{h^{3}\varepsilon c\omega _{R}}{2\chi ^{2}}J_{st}\left[ \omega \left( \chi \omega _{R}^{-1}\right) \right] g\left[ \omega \left( \chi \omega _{R}^{-1}\right) \right] , \label{heat_2}
\end{eqnarray}
where the mean value theorem has been used to approximate the integral, with $0<\chi <1$. Hence, since $\omega \left( \chi \omega _{R}^{-1}\right) =\chi^{-1}\omega _{R}$, Eq. (\ref{heat_2}) reduces to
\begin{equation}
Q=\frac{h^{3}\varepsilon c\omega _{R}}{2\chi ^{2}}J_{st}\left( \chi^{-1}\omega _{R}\right) g\left( \chi ^{-1}\omega _{R}\right) \text{.} \label{heat_3}
\end{equation}
In terms of Planck's distribution, Eq. (\ref{heat_3}) can be rewritten 
\begin{equation}
Q=\frac{\hbar \varepsilon c\omega _{R}}{\chi ^{2}}\frac{g\left( \chi^{-1}\omega _{R}\right) }{\tau ^{\ast }\left( \chi ^{-1}\omega _{R}\right) } \left[ N\left( \chi ^{-1}\omega _{R},T_{1}\right) -N(\chi ^{-1}\omega_{R},T_{2})\right] \text{,} \label{heat_current4}
\end{equation}
while $\tau ^{\ast }(\omega )$ must be determined in a more general way, by scrutinizing the interaction processes between light and bulk material.

For first order in the temperature difference $\Delta T$, 
\begin{equation}
Q=\frac{k_{B}\Delta T\varepsilon c}{\chi }\frac{g\left(\chi^{-1}\omega_{R}\right)} {\tau ^{\ast }\left(\chi ^{-1}\omega _{R}\right)}\left[\frac{\hbar \chi ^{-1}\omega _{R}/2k_{B}T_{0}}{\sinh \left( \hbar \chi ^{-1}\omega_{R}/2k_{B}T_{0}\right) }\right] ^{2} \text{,}  \label{heat_current3}
\end{equation}
with $\Delta T=T_{1}-T_{2}$, and $T_{0}=\left( T_{1}+T_{2}\right) /2$. Thus, it follows that the heat transfer coefficient, the quantity usually measured in experiments and defined by $Q/\Delta T$, is given through 
\begin{equation}
H\left( d,T_{0}\right) =\frac{k_{B}}{\pi^{2}\varepsilon^{2}c^{2}\chi }\frac{\chi ^{-1}\omega _{R}}{\tau ^{\ast }\left(\chi ^{-1}\omega _{R}\right)}\left[\frac{\hbar \chi ^{-1}\omega _{R}/2k_{B}T_{0}}{\sinh\left( \hbar \chi^{-1}\omega _{R}/2k_{B}T_{0}\right) }\right] ^{2} \text{.} \label{heat_transfer2}
\end{equation}

When the mechanism of heat exchange is through \textit{elastic} collisions, which is similar to Rayleigh scattering, it is known that the intensity of radiation is proportional to $\omega^{4}$~\cite{Graebner86}. Therefore, 
\begin{equation}
\tau ^{\ast }(\omega )^{-1}=\tau _{o}^{-1}/4\text{,} \label{eq:conventional}
\end{equation}
where the time scale $\tau _{o}$ is a material-dependent parameter. On the other hand, regarding the \textit{inelastic} contribution to the near-field heat exchange, this is the analogue to the Raman scattering of light. In this case, the distribution of modes presents anomalies which result from states located at a lower energy region~\cite{Chumakov11}. The Raman spectra is fitted using a lognormal function first proposed by Denisov and Rylev~\cite{Denisov90}. This lognormal distribution is a statistical model, which can describe collective motions causing extremely slow structural relaxation, thereby fitting the non-Debye anomalies~\cite{Zanatta10}. Therefore, this accounts for the high nonlinear behavior of the thermal conductance between both materials. For the proposed case, we assume that the density of vibrational states is achieved through the use of a lognormal distribution, which corresponds to 
\begin{equation}
\tau ^{\ast }(\omega )^{-1}=\tau _{o}^{-1}\frac{\omega }{4\sqrt{2\pi } \sigma\omega _{0}}\exp \left[ -\frac{\ln ^{2}\left( \omega /\omega
_{0}\right) }{2\sigma ^{2}}\right] \text{.} \label{eq:D}
\end{equation}
Here, $\omega _{0}$ (characteristic frequency) and $\sigma $ (standard deviation) are two fitting parameters characterizing the lognormal distribution. The lognormal in Eq. (\ref{eq:D}) stems from the existence of a hierarchy of relaxation mechanisms in the material, related to the presence of collective effects. This distribution, results from the fact that the energy of the system consists of a large number of contributions and the application of the central limit theorem of probability theory (see Appendix S1\ref{Appendix}). In view of the properties of the lognormal distribution, it must be noticed that 
\begin{equation}
\int_{-\infty }^{\infty } \frac{1}{\omega \tau^{\ast }(\omega)} d\left( \ln \left( \omega /\omega _{0}\right)\right) =\frac{1}{4\tau _{o}\omega_{0}}\text{,}\label{closure}
\end{equation}
being a kind of closure relation for the relaxation times.

Hence, the excitation of a mode constitutes a photoinduced cooperative phenomenon. It is plausible to assume that the cumulative effect of incident photons ends perturbing the material, thus triggering collective oscillations. In addition, the lognormal accounts for an excess of ways of adsorbing energy by the system with respect to the ways obtained when merely the Debye squared-frequency law describes the relaxation. Therefore, this provides a reasonable description of the non-Debye law~\cite{Shintani08,Chumakov11}, which as in Raman scattering also becomes manifests in radiation problems.

Accordingly, the heat transfer coefficient, Eq. (\ref{heat_transfer2}), results from the addition of the \textit{elastic} and \textit{inelastic}
contributions mentioned above
\begin{equation}
H\left( d,T_{0}\right) =\frac{k_{B}}{\tau_{o}^{\prime }d^{2}}\left[ 1+\frac{(2\pi )^{1/2}\nu c}{\sigma \omega _{0}d}\exp \left\{ -\left[ \frac{\ln\left(2\pi \nu c/\omega _{0}d\right) }{\sqrt{2}\sigma }\right] ^{2}\right\} \right]\left( \frac{h\nu c/2k_{B}T_{0}d}{\sinh \left( h\nu c/2k_{B}T_{0}d\right) }\right) ^{2}\text{,} \label{eq:transfer}
\end{equation}
where $\tau _{o}^{\prime }=\chi ^{3}\tau _{o}$ and $\nu =\varepsilon \chi^{-1}$. In Eq. (\ref{eq:transfer}), the first term inside the square brackets corresponds to the usual contributions found up to now in the current literature~\cite{Mulet02}, taking into account surface phonon-polaritons owing to the presence of evanescent waves close to the interface. As we have mentioned above, the second term takes into account cooperative phenomena, becoming manifest through the existence of collective modes of vibration in the system, which appear in the density of states. Consequently, a more general formulation for the radiative heat transfer problems must come from the superposition of Debye and non-Debye relaxation mechanisms, combining in this way the contributions from the material surfaces as well as the bulk.


\section*{Results and discussion}

The thermal conductance is obtained by integrating the heat transfer coefficient over the surface of a sphere of radius $R$ divided by its area ($A=4\pi R^2$), i.e. $G=(1/A)\int_{0}^{R}H[\tilde{d}(r),T_{0}] 2\pi r dr$. Since the distance $d$ between both surfaces depends on geometrical
characteristics, the local distance between the sphere and the plane surface must be measured through the local radius $r$, assuming the effective distance as $\tilde{d}(r)=d+b+R-\sqrt{R^{2}-r^{2}}$, with $b$ being a surface roughness parameter~\cite{Rousseau09}.

\begin{figure}[!ht]
\begin{center}
\includegraphics[width=4in]{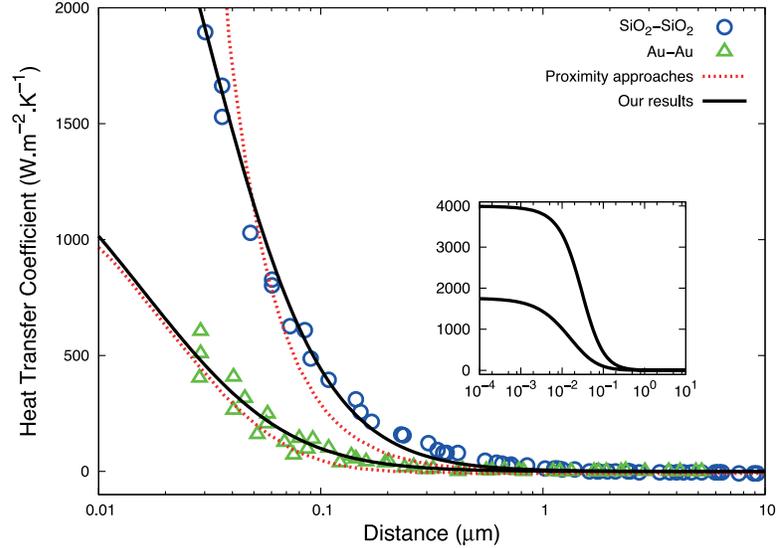}
\end{center}
\caption{ \textbf{Sphere-plate near-field heat transfer coefficients between a gold (or glass) sphere and a gold (or glass) substrate versus gap
distances.} The data are from Ref~\protect\cite{Shen12} for the 50 $\mu$m diameter spheres. The dotted lines are comparisons with the theoretical predictions from the proximity theorem. The inset shows a non-divergent regime as the gap vanishes.} \label{fig:sphere-plate}
\end{figure}

We have calculated numerically the surface average by adjusting the parameters mentioned above to the experimental results obtained in the Ref~\cite{Shen12}, which takes into account only the near-field contribution when decreasing the sphere-plate distance. In Fig. \ref{fig:sphere-plate}, we show the near-field conductance fitting the values of the integral of Eq. (\ref{eq:transfer}) to the data for glass-glass~\cite{Shen09} and gold-gold materials~\cite{Shen12}. For both the material we have used $(\tau_{o}^{\prime })^{-1}=2.1\times 10^{7}$Hz, $\omega_{0}=1.7\times 10^{-13}$ Hz, and $\sigma =6.0\times 10^{18}$. For the glass-glass material we have obtained $\nu =3.2\times 10^{-3}$ and for gold-gold $\nu =7.9\times 10^{-4}$. The near-field heat transfer described using evanescent waves as solutions of classical electrodynamics equations leads to heat flux divergences as the gap vanishes~\cite{Pan00}. However, in our approach this divergence does not
occur, reaching a constant-conductance value as the distance between the nanoparticles decreases (shown in the inset). In essence, this is due to the fact that the time relaxation distribution herein described by lognormal distribution incorporates two effects: (I) Phonon branches in a real
structure affecting the density of states in different frequency regions, similar to actual behavior observed in metals~\cite{Lynn73}; and (II) the
vibrational modes in the bulk material absorbing the energy excess. Hence, since our assumption of a distribution of relaxation times accurately
describes the dynamics of radiative systems at the microscale, we conclude that the FDT in the Callen-Welton formulation~\cite{Callen51} is not
applicable at the nanoscale and must be modified. By going beyond the Debye theory, a way for this generalization is offered here.

\subsection*{Conclusions}
In summary, we have evaluated thermal conductance in the near-field, giving a thermokinetic description of some experiments involving heat radiation through a very narrow gap. Although near-field radiative transfer is a highly complex phenomenon, we have been able to provide a unified and highly accurate explanation of heat exchange processes at the nanoscale. Our theory covers all distances from the far-field up to contact. Since the experiments examined may involve a great variety of nanostructures, our theory possesses a wide scope of applications. The general methodology presented here may also be used in the study of other heat exchange processes such as those occurring in phonon systems and in the analysis of thermal contributions to Casimir forces, even in charge conduction problems in nanosystems.


\section*{Acknowledgments}

This work was supported by MICINN of the Spanish Government, under Grant No. FIS2008-04386, and by CNPq and Funda\c{c}\~{a}o Arauc\'aria of the Brazilian Government. JMR acknowledges financial support from Generalitat de Catalunya under program ICREA Academia.

\bibliographystyle{plain}
\bibliography{template}

\begin{thebibliography}{10}
\providecommand{\url}[1]{\texttt{#1}}
\providecommand{\urlprefix}{URL }
\expandafter\ifx\csname urlstyle\endcsname\relax
  \providecommand{\doi}[1]{doi:\discretionary{}{}{}#1}\else
  \providecommand{\doi}{doi:\discretionary{}{}{}\begingroup
  \urlstyle{rm}\Url}\fi
\providecommand{\bibAnnoteFile}[1]{%
  \IfFileExists{#1}{\begin{quotation}\noindent\textsc{Key:} #1\\
  \textsc{Annotation:}\ \input{#1}\end{quotation}}{}}
\providecommand{\bibAnnote}[2]{%
  \begin{quotation}\noindent\textsc{Key:} #1\\
  \textsc{Annotation:}\ #2\end{quotation}}
\providecommand{\eprint}[2][]{\url{#2}}

\bibitem{Zhang07}
Zhang Z (2007) {N}ano Microscale Heat Transfer.
\newblock New York: {M}cGraw-Hill.
\bibAnnoteFile{Zhang07}

\bibitem{Biehs12}
Morozhenko V (2012) {N}anoscale Radiative Heat Transfer and Its Applications.
  In Infrared Radiation.
\newblock {I}nTech.
\bibAnnoteFile{Biehs12}

\bibitem{Basu07}
S~Basu YBC, Zhang ZM (2007) {I}nt {J} {E}nergy {R}es 31: 689.
\bibAnnoteFile{Basu07}

\bibitem{Polder71}
Polder D, van Hove M (1971) {P}hys {R}ev {B} 4: 3303.
\bibAnnoteFile{Polder71}

\bibitem{Rytov59}
Rytov SM (1959) {T}heory of electric fluctuations and thermal radiation.
\newblock Bedford: {AFCRC-TR}. {A}ir {F}orce {C}ambridge {R}esearch {C}enter,
  {A}ir {R}esearch and {D}evelopment {C}ommand, {U.S.} {A}ir {F}orce.
\bibAnnoteFile{Rytov59}

\bibitem{Mulet02}
Mulet JP, Joulain K, Carminati R, Greffet JJ (2002) {M}icroscale {T}hermophys
  {E}ng 6: 209.
\bibAnnoteFile{Mulet02}

\bibitem{Rousseau09}
Rousseau E, Siria A, Jourdan G, Volz S, Comin F, et~al. (2009) {N}ature
  {P}hoton 3: 514.
\bibAnnoteFile{Rousseau09}

\bibitem{Shen09}
Shen S, Narayanaswamy A, Chen G (2009) {N}ano {L}ett 9: 2909.
\bibAnnoteFile{Shen09}

\bibitem{Shen12}
Shen S, Mavrokefalos A, Sambegoro P, Chen G (2012) {A}ppl {P}hys {L}ett 100:
  233114.
\bibAnnoteFile{Shen12}

\bibitem{Pendry99}
Pendry JB (1999) {J} {P}hys: {C}ondens {M}atter 11: 6621.
\bibAnnoteFile{Pendry99}

\bibitem{Biehs10}
Biehs SA, Rousseau E, Greffet JJ (2010) {P}hys {R}ev {L}ett 105: 234301.
\bibAnnoteFile{Biehs10}

\bibitem{Sasihithlu11}
Sasihithlu K, Narayanaswamy A (2011) {P}hys {R}ev {B} 83: 161406(R).
\bibAnnoteFile{Sasihithlu11}

\bibitem{Domingues05}
Domingues G, Volz S, Joulain K, Greffet JJ (2005) {P}hys {R}ev {L}ett 94:
  085901.
\bibAnnoteFile{Domingues05}

\bibitem{Forster48}
F\"{o}rster T (1948) {A}nn {P}hys 2: 55.
\bibAnnoteFile{Forster48}

\bibitem{Madrid08}
P\'{e}rez-Madrid A, Rub\'{\i} JM, Lapas LC (2008) {P}hys {R}ev {B} 77: 155417.
\bibAnnoteFile{Madrid08}

\bibitem{Madrid09}
P\'{e}rez-Madrid A, Lapas LC, Rub\'{\i} JM (2009) {P}hys {R}ev {L}ett 103:
  048301.
\bibAnnoteFile{Madrid09}

\bibitem{Madrid10}
P\'{e}rez-Madrid A, Rub\'{\i} JM, Lapas LC (2010) {J} {N}on-{E}quilib
  {T}hermodyn 35: 279.
\bibAnnoteFile{Madrid10}

\bibitem{deGroot84}
de~Groot SR, Mazur P (1984) {N}onequilibrium thermodynamics.
\newblock {N}ew {Y}ork: {D}over.
\bibAnnoteFile{deGroot84}

\bibitem{Denisov90}
Denisov YV, Rylev AP (1990) {JETP} {L}ett 52: 411.
\bibAnnoteFile{Denisov90}

\bibitem{Shintani08}
Shintani H, Tanaka H (2008) {N}ature 7: 870.
\bibAnnoteFile{Shintani08}

\bibitem{Chumakov11}
Chumakov A, Monaco G, Monaco A, Crichton WA, Bosak A, et~al. (2011) {P}hys
  {R}ev {L}ett 106: 225501.
\bibAnnoteFile{Chumakov11}

\bibitem{Planck91}
Planck M (1991) {T}he theory of heat radiation.
\newblock {N}ew {Y}ork: {D}over.
\bibAnnoteFile{Planck91}

\bibitem{Yager36}
Yager W (1936) {P}hysics 7: 434.
\bibAnnoteFile{Yager36}

\bibitem{Debye29}
Debye P (1929) {P}olar molecules.
\newblock {N}ew {Y}ork: Chemical Catalog Co.
\bibAnnoteFile{Debye29}

\bibitem{Callen51}
Callen H, Welton T (1951) {P}hys Rev 83: 34.
\bibAnnoteFile{Callen51}

\bibitem{Graebner86}
Graebner JE, Golding B, Allen LC (1986) {P}hys {R}ev {B} 34: 5696.
\bibAnnoteFile{Graebner86}

\bibitem{Zanatta10}
Zanatta M, Baldi G, Caponi S, Fontana A, Gilioli E, et~al. (2010) {P}hys {R}ev
  {B} 81: 212201.
\bibAnnoteFile{Zanatta10}

\bibitem{Pan00}
Pan JL (2000) {O}pt {L}ett 25: 369.
\bibAnnoteFile{Pan00}

\bibitem{Lynn73}
Lynn JW, Smith HG, Nicklow RM (1973) {P}hys {R}ev {B} 8: 3493.
\bibAnnoteFile{Lynn73}

\end{thebibliography}

\section*{Figure Legends}

\section*{Appendix S1: Lognormal distribution}\label{Appendix}

This Appendix is devoted to the derivation of the expression of the lognormal distribution corresponding to Eq. (18). The energy of the system results from the combined effect of a large number of independent inputs $e_{\alpha }$, $e_{1}+...+e_{n}$. Let us assume that the effect on the energy of the input $e_{\nu }$ is proportional to $e_{\nu }$ and to the cumulative effect $E_{\nu}$ of the $\nu -1$ previous inputs, $E_{\nu +1}=E_{\nu }+e_{\nu }E_{\nu }$. Whence 
\begin{equation}
e_{1}+...+e_{n}=\sum_{1}^{n}\frac{E_{\nu +1}-E_{\nu }}{E_{\nu }}\simeq \frac{1}{k}\int_{E_{1}}^{E}\frac{d\epsilon }{\epsilon }=\frac{1}{k}\log \frac{E}{E_{1}}
\end{equation}
For large $n$, the distribution of the sum is given by a Gaussian, according to the central limit theorem
\begin{equation}
\rho \left( \omega \right) \sim \exp \left[ -\left( \frac{1}{k}\log \frac{\omega }{\omega _{1}}\right) ^{2}/2\sigma ^{2}\right]
\end{equation}
where $E=\hbar \omega $. Equivalently
\begin{equation}
\rho \left( \omega \right) d\omega =\frac{k\omega _{1}}{\sqrt{2\pi }\sigma\omega }\exp \left[ -\left( \frac{1}{k}\log \frac{\omega }{\omega _{1}}\right) ^{2}/2\sigma ^{2}\right] d\omega \text{.}
\end{equation}
This distribution depends on two empirical parameters $\sigma $ and $\omega
_{1}$.
\end{document}